# METROLOGY INFLUENCE ON THE CUTTING MODELISATION

## (DE L'INFLUENCE DE LA MÉTROLOGIE SUR LA MODÉLISATION DE LA COUPE)

Olivier CAHUC, Alain GERARD

*Abstract: High speed machining has been improved thanks to considerable advancement on the tools (optimum geometry, harder materials), on machined materials (increased workability and machining capacity for harder workpieces) and finally on the machines (higher accuracy and power at the high speeds, performances of the numerical control system).*

*However at such loading velocities, the cutting process generates high strain and high-speed strain which cause complex, various and irreversible phenomena in plasticity.*

*These phenomena are comprehended through the complete measurement of the mechanical actions using a six-component dynamometer and flux and temperatures measurements at the tip of the cutting tool. Balanced energy assessments are the starting points of our reflection on the machining modelling.*

*The modelling of these phenomena and the material behaviour under this type of loading requires a suitable theoretical approach. The main points of the strain gradient theory are developed. A theoretical behaviour law adapted to the cutting phenomena is then expressed.*

## 1. INTRODUCTION

**Evolutions and revolution of the scientific approach to the cutting process**

In the highly competitive industrial sector of machining, it has now become necessary to control the production costs. They are primarily due to the duration of machining, the type of machine used, the consumption of tools and lubricants. For a number of years, technological advances such as tool materials or numerical control machines have enabled industrialists to maintain a linear reduction of the production costs without having to resort to scientific methods. A few series of tests could thus easily solve a particular problem and give satisfaction to the customer via a material or a specific tool geometry. Currently, this tendency is undergoing strong changes; and moreover industrialists are offered few alternative solutions outside High Speed Machining, which is seen as a radical technological step by many of them. As a matter of fact, the improvements in performance in the last ten years relate more to increased performances in data processing than to an actual optimisation of the tool-matter couples.

Nevertheless, the computing capacities of C.A.M. software or of the numerical controls do not permit to exceed the physical limits imposed by a non-optimised chip formation. Today, the problem arises in many advanced machining industries, which have already reached these limits and tool or machine manufacturers are looking for a clear solution to meet the strong demand for increased productivity. In order to satisfy the industry's expectations, it seems essential to deal with the problem as a whole. The modelling of the cutting process is a response to the problems raised by this technology. However, this requires a comprehensive understanding of the phenomena. In this respect, the difficulty of apprehending the global geometry of the process can be better highlighted by passing from a 3D to a plane representation (according to A view on Fig. 1) to identify the most characteristic zones of machining. In fact, the majority of the scientific community dealing with the subject is aware of the importance of the three-dimensional analysis. To be convinced of the non-relevance of the two-dimensional - or assimilated - aspect of these starting assumptions, it is just necessary to carry out tests on machine and to observe the chip formation (Fig. 2).

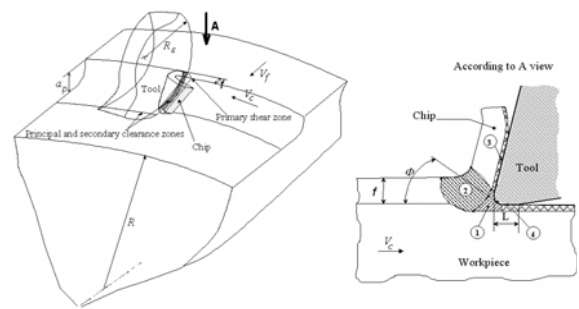

**Fig. 1.** Main machining zones

Experimental results of turning [Toulouse, 1998] were obtained at the LMP (Research Laboratory of the University Bordeaux 1) using the six component dynamometer [Couétard, 1993], which allowed COUÉTARD [Couétard, 2000] to inspect the cutting process as a truly three-dimensional phenomenon. For this purpose, he used the whole range of the mechanic's tools and he focused more particularly on the action of a body on another body through a force vector and a moment vector.

We strongly advise users of the cutting process to treat it in a three-dimensional space in order to avoid making assumptions, which might reduce the study to an orthogonal (or oblique) case, which would fail to reflect the real machining process in turning, drilling or milling. Such assumptions are too restrictive in high speed or very high speed machining. Cutting speeds are increasing due to the development of tools and materials. Thus, the very high deformation velocities in the cutting process involve high strain velocities and strain [Moufski et al, 2000]. The behaviour of the materials under such conditions is still unknown and it seems possible to obtain a better expression of the material load by improving the behaviour laws of the material. The friction and temperature (or flux) conditions to the interfaces between the tool and the chip are not easy to characterize and to model. The older mechanical models are too simplistic to explain and predict the physical phenomena occurring in the cutting process. Therefore, more complete analytical or numerical models must be developed for the cutting process to be representative of fit today's industrial reality. Then, at a time when many companies are adopting HSM, it is necessary to change the two-dimensional models for a three-dimensional space in which the tool-chip contact will not be regarded as a punctual contact.

Among the current lacks of different existing models, one could mention:
- the necessity to take into account the 3D aspect of the mechanical actions exerted on the material,
- the necessity to produce a better analysis of the thermal aspects, their measurement and their modelling,
- the difficulty to develop experimental devices recreating the phenomena occurring at the tool-chip contact and to analyse it,
- the necessary development of new more complete behaviour laws adapted to the complex transformation of material.

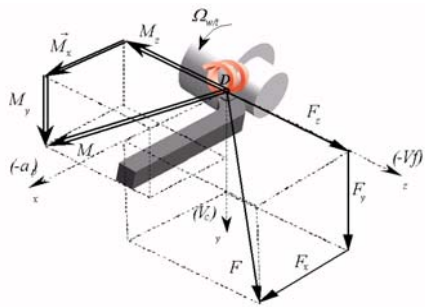

**Fig. 2.** Three-dimensional cut in turning (kinematics and cutting actions)

## 2. MOMENTS AT THE TIP OF THE TOOL

The synthesis of experimental statements in turning has shown the presence of the six mechanical actions between the workpiece and the tool [Toulouse et al, 1997; Cahuc et al, 2001]. The existence of two (orthogonal cut) or even three (oblique cut) forces components is beyond question. Conversely, the presence of a vector moment directly applied to the tip of the tool and not directly resulting from the preceding force vector is much more debatable. This concept is goes against all the theories based on the only forces vector with the mechanical effects on the stress and strain states in the various zones identified. For the first time, such an approach has made it possible to obtain a balanced energy assessment, i.e. an initial power input into a machine tool has got thoroughly transformed into mechanical spindle power while being degraded into thermal energy through the cutting process [Cahuc et al, 2001].

### 2.2. Experimental devices

The experimental devices used allow to measure the 6 components of the mechanical actions and the thermal flux on the tip of the tool during machining. The 6 components of the mechanical actions are measured with a 6 component dynamometer [Couétard, 1993] (Fig. 3).

The principles and the procedures of calibration can be found in COUETARD's Thesis [Couétard, 2000]. The dynamometer accuracy is about $\pm$ *8N* for forces and $\pm$ *0.5Nm* for moments in this loading case. Thus, the six contact actions in real time during the cutting process should be obtained.

### 2.3. Examples of experimental mechanical results

*2.3.1. Turning mechanical results*

An example of result obtained during an operation of hard turning on a cemented steel (65HRC hardness) is given. The tool insert used is a CBN TNMA 16 04 08 from SANDVIK COROMANT (Vc=173 m.mn$^{-1}$ - f=0.1 mm/rev. - ap=0.185 mm). The vector moment is expressed at the tip of the tool (Fig. 4).

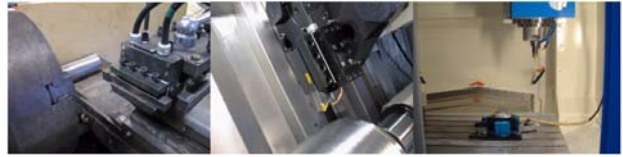

**Fig. 3.** Six component turning dynamometer (on conventional and CNC lathe) and milling one

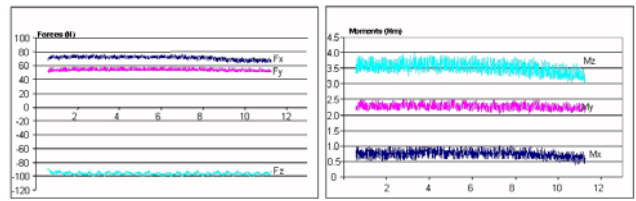

**Fig. 4.** Measurement of the forces Fx, Fy, Fz and of the moments Mx, My, Mz during an operation of turning.

*2.3.2. Milling mechanical results*

High-speed milling tests (N=17000rpm$^{-1}$-Vf=4080 mm.min$^{-1}$-ap=6mm-ar=1mm) were carried out on an aluminum alloy piece with a double-edged milling tool under the configuration given on Fig. 5. The mechanical actions measured during this machining are given in Fig. 6.

## 2.4. Analysis of the mechanical experimental results

*2.4.1 Considerations on energy*

The presence of all the components has a direct effect on the total power consumed by the process. Indeed, this mechanical power can be expressed with the static torsor T and the kinematics torsor C in a reference axis system related to the tool. Its origin D is a point located, by assumption, in the middle of the cutting edge.

$$P_m = \{T_{Tool \to Workpiece}\}_{D,x,y,z} \cdot \{C_{Workpiece \to Tool}\}_{D,x,y,z}, \quad (1)$$
$$P_m = F_X V_X + F_Y V_Y + F_Z V_Z + M_X \Omega_X + M_Y \Omega_Y + M_Z \Omega_Z.$$

During a longitudinal turning test in the direction -$\vec{z}$ (Fig. 2), the kinematics torsor has 3 non-null components $V_Y$, $V_Z$ et $\Omega_Z$. The total mechanical power can be expressed as

$$P_m = -F_Y V_Y - F_Z V_Z - M_Z \Omega_Z, \quad (2)$$

where $V_Y$ and $V_Z$ are respectively the cutting speed and feed velocity expressed in m.s$^{-1}$ and $\Omega_Z$ the spindle speed expressed in rad.s$^{-1}$.

For the tests carried out, the feed velocity used being very weak, the term $F_Z V_Z$ is approximately only 2% of the total consumed power. Then, it can be neglected.

In practice, many users take into account the first term only ($P_m = F_Y \cdot V_Y$) as they assume that the tool-chip contact is an 'almost-punctual' contact.

The torsor between the two bodies is then summarized with a two-component force vector.

It must be noted that while this assumption was acceptable in 1945 – it served as a reference for reference models [Merchant, 1945a; 1945b] – it is much too simplistic today.

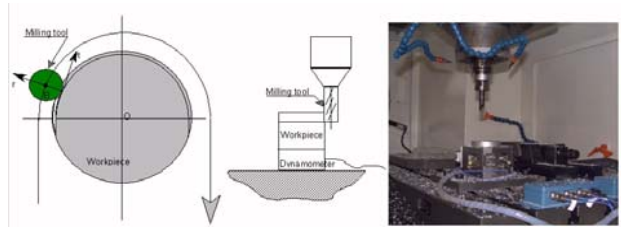

**Fig. 5.** High Speed Milling      Experimental device

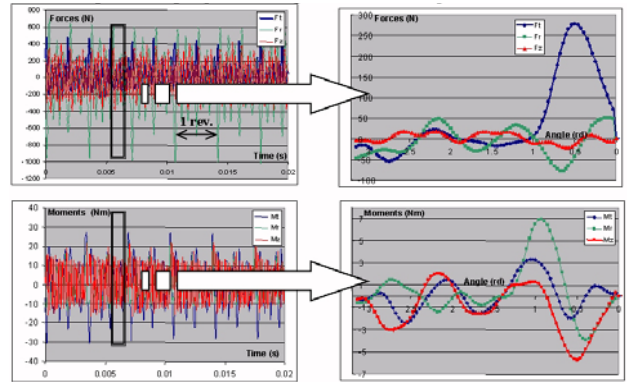

**Fig. 6.** Extracted acquisition file - Results for an half revolution

For the preceding hard turning operations, we have shown [Darnis et al, 2000; Couétard et al, 2001] that the part power consumed by the term $F_Y V_c$ accounts for only 60% of the total power (Table 1).

| $F_y V_c$ | $F_z V_f$ | $M_z \Omega_z$ | $P_m$ |
|---|---|---|---|
| 277.5 ± 23.5 W | –0.07 ± 0.01 W | 190 ± 42 W | 467 ± 65 W |
| 60% | ; 0% | 40% | 100% |

**Table 1.** Mechanical power in hard turning

Similarly, the results obtained during the high-speed milling tests, show how important it is to integrate all the components of the cutting actions torsor to obtain a better analysis of the phenomena involved (Table 2).

| $F_y V_c$ | $M_z \Omega_z$ | $P_m$ |
|---|---|---|
| 3966 W | 10254 W | 14220 W |
| 27.9% | 72.1% | 100 % |

**Table 2.** Mechanical power in high speed milling

### 2.5. Conclusion

This various experimental work shows that the moments at the tip of the tool are real and that their presence has an influence on the total cutting power consumed and the geometry of the chip, its evacuation, the machined surface quality, the tool, and its wear. The reliable information we obtained on the mechanical actions led us to search for the experimental fluxes and temperatures generated by the process in order to chart its energy consumption.

## 3. TEMPERATURE MEASUREMENT DEVICE AT THE TIP OF THE TOOL

The presence of the chip in front of the device limits the use of any existing commercial solution (pyrometer, infra-red camera...).

In order to link the flux on the cutting edge with the temperature, a model of the heat transfer in the tool has been realized with an inverse method using a tool instrumentation (Fig. 7) to measure the temperature in one or more points far away from the cutting edge (Fig. 9).

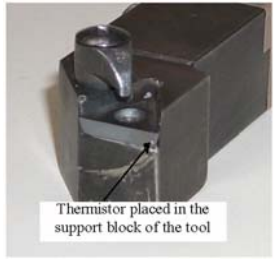

**Fig. 7.** Installation of the sensor on the tool

Phase ①: Tool characterization. The parameters of the non-integer model of the tool behaviour are estimated from the measurements on an experimental device, which reproduce the thermal conditions of machining on the cutting edge of the tool.

Phase ②: The procedure of inversion is carried out starting from the model found with the temperature characterization and tool sensor measurements.

Phase ③: The evaluation of fluxes and tool temperature is carried out during machining.

The complete work [Battaglia et al, 1999; Battaglia et al, 2000a; Battaglia et al, 2000b; Puigsegur et al, 2001; Battaglia, 2002; Puigsegur, 2002] allows to look further into the methods used.

### 3.1. Examples of experimental results

#### 3.1.1. Thermal measurement results in turning

It is possible to estimate the flux and the average temperature on the cutting edge of the tool by using the inversion procedure and the temperature measurement.

The signals measured at the sensor and calculated by the model during the machining test are presented in .

Finally, this experimental work on the thermal aspect of machining has allowed to work out a new experimental methodology in order to measure the temperature and the flux at the tip of the tool. Other works have allowed us to obtain the temperature on the workpiece surface during machining [Battaglia et al, 2005].

This methodology allows to improve existing models [Komanduri et al, 2000, 2001a; 2001b] of heat transfers in the workpiece, the chip and the tool without having to consider the contact resistance value between the tool and the chip.

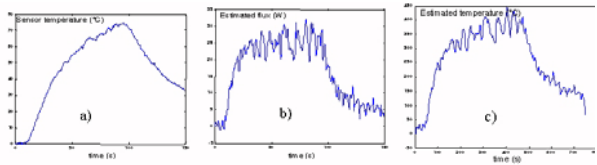

**Fig. 8.** a) Temperature measured at the sensor b) Flux and c) average temperature estimated at the tip of the tool

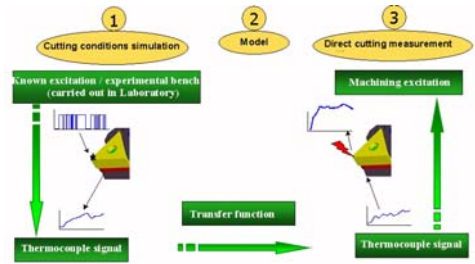

**Fig. 9.** Temperature measurement at the tip of the tool

## 4. MECHANICAL/ THERMAL ENERGY BALANCE

To conclude on the experimental aspect, some turning tests allow to confront the three consumed powers, $P_E$ brought by the pin of the turn, $P_m$ established using the six-component measurement of the cutting actions and finally $P_{th}$ calculated with the temperature measurement at the tip of the tool. For the first time a mechanical/thermal energy assessment [Cahuc et al, 2001] is obtained with these three powers.

Two series of tests are carried out under traditional turning conditions on a CC35 steel with a TNMM 16 04 08 insert (first configuration) and under hard turning conditions on a cemented steel (65 HRC hardness) with a TNMM 16 04 08 CBN insert (second configuration).

The results obtained for the three different powers by the various experimental methods described in the preceding paragraphs appear in Table 3.

| First configuration | | | Second configuration | | |
|---|---|---|---|---|---|
| $P_m$ | $P_{th}$ | $P_E$ | $P_m$ | $P_{th}$ | $P_E$ |
| 1194W | 1283W | 1139W | 467W | 477W | 439W |

**Table 3.** Comparison between the three powers: mechanical Pm, thermal P th and provided to the spindle $P_E$

The small variations (since *5%* to *6.5%* between the power provided to the spindle $P_E$ and the mechanical power $P_m$ and *7.5%* to *8.6%* between $P_E$ and the thermal power $P_{th}$) point to [Cahuc et al, 2001] a complete energy assessment in the case of turning.

A direct consequence of these results is the necessity to take into account the complete mechanical tool/material aspect to study and to model the cutting phenomena.

## 5. EXPERIMENTAL PART CONCLUSION

The mechanical and thermal experimental devices developed make it possible to obtain reliable results, which cannot leave any more of doubt for the presence of a complete action torsor at the tip of the tool during machining. This torsor alone does not bring a direct response to the problems of process optimisation, which posed to the manufacturers. But analysing its evolution during machining or comparing it with 2 machining processes allows, together with other parameters such as tool wear or roughness, to improve comprehension of the phenomena. It also allows to obtain substantial productivity gains for example.

These various aspects have opened new investigation fields for analysing and modelizing these phenomena correctly. This work allows to understand that the phenomena identified cannot be modelled with the existing behavioural approaches and that it is necessary to develop adapted behaviour laws.

## 6. THREE-DIMENSIONAL SEMI-ANALYTICAL THERMO MECHANICAL CUTTING MODEL

### 6.1. Introduction

The modelling of the complex phenomena during a machining operation has to be realized with the previous results. The first model developed by TOULOUSE [Toulouse, 1998] and the following models which evolved from it always went in the direction of the energy aspect of the phenomena.

Then, the main characteristic of the model is the prediction of the presence of forces and moments at the tip of the tools during chip formation, in order to minimize the power consumed by the cutting process.

### 6.2. Thermomecanical model description

The three-dimensional aspect of the initial model aims to break from the former plane models and to deal with the sets of the requested zones: two shear zones (primary ② and secondary ③) as well as the two rake zones (primary ④ and secondary ⑤)(cf.Fig. 1). The zone ① is the common zone with the four preceding ones and allows their connection.

The principal assumptions relative to this model are the permanent rate of flow, homogeneous and isotropic medium with Von Mises plastic threshold, stationary and incompressible plastic flow, constant shear stress along the average primary shearing line.

The principal developments and results of this initial model are presented in [Toulouse, 1998; Cahuc et al, 2001].

The equilibrium of the tool and the chip is expressed to determine the cutting actions applied to them (Fig.10).

The tool/workpiece contact zone is considered as a continuous surface (Fig.10). The cutting and rake faces are included in this zone. The connection of these surfaces is made by the tool radius $r_\varepsilon$ and the edge acuity $R$. This approach is in accordance with the modern cutting tools such as insert tools generally having large radius. Moreover, it allows to define the tool/chip/workpiece contact interface in a more complete way.

The aim is now to write an analytical thermal model of the phenomena occurring during an operation of turning in order to integrate it into the previous model.

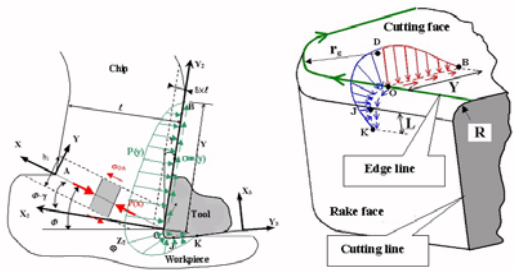

**Fig.10.** Loading representation and geometrical description

The existing models use empirical results [Boothroyd, 1963] for the primary shear zone and these do not always permit to define the divisions of flux in the various zones [Tay et al, 1976].

It is assumed that the mechanical energy coming from plasticity, viscous dissipation of machined material and frictions of the tool on the workpiece and on the chip is degraded into thermal energy [Cahuc et al, 2001]. The physical phenomena in the cutting zone, contribute to generating a heat flux in the tool, the chip and the workpiece. The connection between the variations of these thermo-mechanical phenomena to the cutting parameters is fundamental for a total control of the process.

Our purpose was to characterize the thermal transfers in the cutting zones. This study was undertaken with analytical models, and identification of not-integer systems of opposite problem resolution.

The modelling of the thermal transfers in the workpiece and the chip was carried out analytically. It made it possible to give a partition of the heat flux and to establish a thermal mapping of the workpiece and the chip.

The modelling of the thermal transfers in the cutting tool was carried out differently. The average temperature is measured by the sensor placed in the tool (Fig. 7). The estimate of the thermal requests imposed on the tool are then carried out in real time, by resolution of the thermal inverse problem [Battaglia et al, 2000a; Battaglia et al, 2000b; Battaglia, 2002]. This method integrates all the tool complexity (several parts and thermal contact resistance unknown factor) through a specific experimental device and a specific calibration.

In the primary shear zone $(OA)$, the influence of the strain and strain velocity is predominating over that of the temperature. We chose to use the modified JOHNSON-COOK law [Joyot, 1994]. It does not take into account the temperature dependence of the mechanical behaviour of material for the material flow characterization in this zone:

$$\sigma_{OA} = \left(A + B\varepsilon_{OA}^{n}\right)\left[1 + C\ln\left(\frac{\dot{\varepsilon}_{OA}}{\dot{\varepsilon}_0}\right)\right]. \qquad (3)$$

Concerning the secondary shear zone, the shear stress along $(OB)$ is calculated with a complete Johnson Cook behaviour law:

$$\sigma_{OB} = \left(A + B\varepsilon_{OB}^{n}\right)\left[1 + C\ln\left(\frac{\dot{\varepsilon}_{OB}}{\dot{\varepsilon}_0}\right)\right]\left[1 - \left(\frac{T_{OB} - T_{amb}}{T_{fus} - T_{amb}}\right)^m\right] (4)$$

A study [Puigsegur, 2002] then allowed us to show the sensitivity of the previous various source terms to the JOHNSON COOK law parameters. It is important to note that the combination of approximations on all the parameters of the behaviour law and friction law, can lead us to erroneous values of the sources terms.

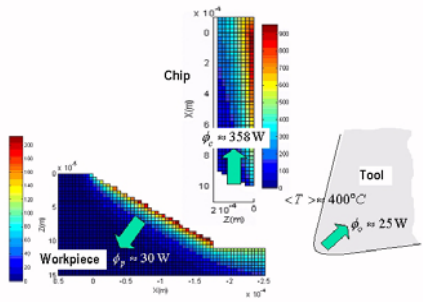

**Figure 11.** Distribution of flux and thermal cartography of the cutting

The thermal balance sheet in the cutting zone is thus carried out zone by zone:
- we have used the work results of KOMANDURI and HOU [Komanduri et al, 2000, 2001a; 2001b] in the primary shear zone,
- in the secondary shear zone, the model of the chip developed by KOMANDURI and HOU has been used
- the precedent model has been extended to the clearance zone,

The results obtained [Puigsegur, 2002] enable us to present an example of a complete heat assessment of the cutting zone on Figure 11 (i.e. the distribution of the flux in the tool, the workpiece and the chip and the thermal mapping of the workpiece, of the chip as well as the estimated average temperature on the cutting edge of the tool).

## 7. ANALYSIS AND DISCUSSION ON THE VALIDITY OF THE BEHAVIOUR LAWS USED CLASSICALLY IN CUTTING MODELS

**Second gradient theory – New behaviour law forms for the secondary shear zone:**

Our most recent work [Cahuc, 2005] concerned the validity of the behaviour laws in these zones and in particular in the more complex secondary shear zone. It was selected because the present phenomena were more easily reproduced on an experimental device independent from the machining operation. Thus, the material behaviour could be identified through new forms of behaviour laws.

The moment can be due to the effect of a sub-surface strain (Figure 12) with strong gradient.

The most adequate generalized theory is the second gradient theory. It is a generalization of the classical theory of the continuous media [Germain, 1973a] and, the stress couple theory can be developed from this theory [Germain, 1973b] with some specific considerations. The description of the movement allows to introduce the material strain as in the classical theory of the continuous media, as well as the strain gradient (Table 4).

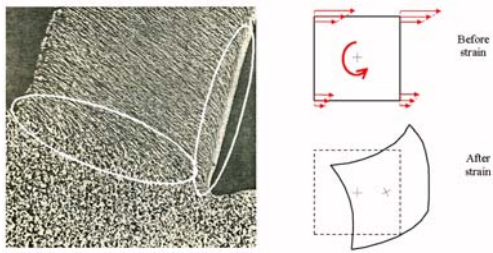

**Figure 12.** Micrograph of the strain gradient zones and $2^d$ gradient modelling

The movement description becomes then more precise and the strain gradients are in agreement with the rotation strain phenomena modelling. This method was developed in the 1960s by MINDLIN [Mindlin et al, 1968], TOUPIN [Toupin, 1962] and GERMAIN [Germain, 1973b]. These works defined the bases of the second gradient theory, but only in the case of linear elasticity. The case of plasticity was not treated until the works of FLECK and HUTCHINSON [Fleck et al, 1997] and finally the case of visco-plasticity was treated by GURTIN [Gurtin, 2003].

The new formulations obtained for the boundary condition and equilibrium equations indicate the elements, which must appear in the behaviour laws.

These laws are defined by the following equations:

$$\dot{\varepsilon}_{ij}^p = \frac{3}{2} H(f) g'(\Sigma_{eq}) \frac{\langle \dot{\Sigma}_{eq} \rangle}{\Sigma_{eq}} \sigma'_{ij}, \quad (5)$$

$$\dot{\eta}_{ijk}^p = H(f) g'(\Sigma_{eq}) \frac{\langle \dot{\Sigma}_{eq} \rangle}{\Sigma_{eq}} \sum_{I=1}^{3} \left[ 1_I^{-2} \mu'^{(I)}_{ijk} \right]. \quad (6)$$

The overall effective stress tensor $\Sigma_{eq}$ is then defined by:

$$\Sigma_{eq} = \sqrt{\overline{\sigma}^2_{eq} + \overline{\overline{\mu}}^2_{eq} + \sum_{I=1}^{3}\left[1_I^{-2}\overline{\overline{\mu}}^{2\,I}_{eq}\right]} \quad (7)$$

$$= \sqrt{3.(\sigma_{zy})^2 + \frac{3}{2}.(\mu_{zx})^2 + \frac{1}{1_c}\left(\frac{399}{160}(\mu_{yzz})^2 + \frac{229}{90}(\mu_{yyz})^2\right)}.$$

The definition of the form of the work hardening function $g'(\Sigma_{eq})$ characterizing the plasticity state in relations (5) and (6) requires the evaluation of the field of strain velocity.

| Classical theory (first gradient) |
|---|
| $\mathcal{P}_i + \mathcal{P}_c + \mathcal{P}_d = \mathcal{P}_a$ |
| $\mathcal{P}_i = -\int_{\mathcal{D}} \sigma_{ij} \dot{\varepsilon}_{ij} d\upsilon$ |
| $\mathcal{P}_i = \int_{\mathcal{D}} \sigma_{ij,j} U_i d\upsilon + \int_{\partial\mathcal{D}} \sigma_{ij} n_j U_i dS$ |
| $\mathcal{P}_c = \int_{\partial\mathcal{D}} T_i U_i dS$ |
| Second gradient theory |
| $\mathcal{P}_i + \mathcal{P}_c + \mathcal{P}_d = \mathcal{P}_a$ |
| $\mathcal{P}_i = -\int_{\mathcal{D}} \left( \sigma_{ij} \dot{\varepsilon}_{ij} + \mu_{ijk} \dot{\eta}_{ijk} \right) d\upsilon$ |
| $\mathcal{P}_i = \int_{\mathcal{D}} \mathcal{F}_i U_i d\upsilon + \int_{\partial\mathcal{D}} \left( T_i U_i + \mathcal{M}_i \omega_i + \mathcal{N}_i D_{\underline{nn}} \right) dS + \int_{\Gamma} \mathcal{R}_i U_i ds$ |
| $\mathcal{P}_c = \int_{\partial\mathcal{D}} \left( T_i U_i + M_i \omega_i \right) dS + \int_{\Gamma} R_i \left( DU_i \right) ds$ |

**Table 4.** *First and the second gradient theories*

A first approach of measurement of this field within the material was elaborated.

The micrograph (Figure 12) reveals the trajectory of a material particle. From this trajectory, it is possible to determine the evolution of the displacement field rates and indirectly the field of strain velocity. The observation is carried out after loading, which does not make it possible to estimate the duration of displacements during the test: the current test duration being 10 seconds, it is impossible to determine if the strain state was obtained at the end of 2 seconds, 5 seconds or 10 seconds. A complementary study must be realised to remove this ambiguity and to obtain more information on the evolution of the strain velocity field.

## 8. GENERAL CONCLUSION

The machining process introduce complex phenomena during chip formation, relating to various fields such as deformable solid mechanics or thermal phenomena which make it difficult to control the energy parameters. The measurements carried out using a six-component dynamometer allow a better understanding of these phenomena. They reveal the presence of moments at the tip of the tool, which the traditional cutting models fail to evaluate.

A three-dimensional cutting modelling integrating this concept of moment and based on the theory of the second gradient theory was adopted to describe the cutting phenomena occurring in the secondary shear zone.

Therefore, the principal developments of the second gradient theory were developed. Works resulting from the literature [Germain, 1973a; 1973b] present a method theory where all the terms which might be encountered are defined. Recent works of FLECK and HUTCHINSON [Fleck et al, 1997] have made it possible to establish the general form of the behaviour laws using the second gradient theory in the case of plasticity. Actually, works dealing with the cases of high strain and high strain velocity do not exist, as yet.

The theoretical elements were presented but the finality of this work is to provide behaviour laws adapted to the cutting phenomena in the two shear zones. The analytical or numerical resolution of the presented equations has not been carried out yet. This resolution will provide the evolution of the stresses and stress gradient fields applied to the material. The microscopic observations carried out after loading as well as an analysis of the strain state will provide the strain and the strain gradient fields. Then, the behaviour law defines the connection between stresses and strain as well as their gradients. Complementary experiments must be carried out in order to determine the evolution of the strain and strain gradient fields completely.

Significant experimental and theoretical works remain to be carried out, first to determine the values of the coefficients of the behaviour law in the secondary shear zone and also to show that this theory can also find its application in primary shear zone.

O. CAHUC, Maître de Conférences HDR , Université Bordeaux 1, Laboratoire de Mécanique Physique UMR CNRS 5469,
olivier.cahuc@u-bordeaux1.fr
Phone: +33 (0)540 008 789
Fax: +33 (0)540 006 964

Alain GERARD, Professeur des universités, Université Bordeaux 1, Laboratoire de Mécanique Physique UMR CNRS 5469,
alain.gerard@u-bordeaux1.fr
Phone: +33 (0)540 006 223
Fax: +33 (0)540 006 964